\documentstyle[12pt]{article}
\begin{document}
\setlength{\unitlength}{1mm}
\newcommand{\te}{\theta}
\newcommand{\bee}{\begin{equation}}
\newcommand{\ene}{\end{equation}}
\newcommand{\bea}{\begin{eqnarray}}
\newcommand{\ena}{\end{eqnarray}}
\newcommand{\tra}{\triangle\theta_}

{\hfill Universit\'a di Napoli Preprint DSF-22/95} \vspace*{1cm} \\

\begin{center}
{\Large\bf Full Phase-Space Analysis of Particle Beam }\\
\end{center}
\begin{center}
{\Large\bf Transport in the Thermal Wave Model}
\end{center}

\bigskip\bigskip

\begin{center}
{{\bf R.~Fedele}$^{1,2}$, {\bf F.~Galluccio}$^{2}$, {\bf V.I.~Man'ko}$^{1,3}$
and {\bf G.~Miele}$^{1,2}$}
\end{center}

\bigskip

\noindent
{\it $^{1}$ Dipartimento di Scienze Fisiche, Universit\`a di Napoli
''Federico II'', Mostra D'Oltremare Pad. 20, I-80125 Napoli, Italy}\\

\noindent
{\it $^{2}$ Istituto Nazionale di Fisica Nucleare, Sezione di Napoli,
Mostra D'Oltremare Pad. 20, I-80125 Napoli, Italy}\\

\noindent
{\it $^{3}$ Lebedev Physics Institute, 53 Leninsky Prospect, 117924
Moscow, Russia}

\bigskip\bigskip\bigskip

\begin{abstract}
Within  the Thermal Wave Model framework a comparison among Wigner
function, Husimi function, and the phase-space distribution given by a
particle tracking code is made for a particle beam travelling through
a linear lens with small aberrations. The results show that the
quantum-like approach seems to be very promising.
\end{abstract}

\vspace{1.cm}
\centerline{\it to appear in Phys. Lett. A}

\vspace{3.cm}

{\footnotesize
\begin{itemize}
\item[] E-mail:\\
	Fedele@axpna1.na.infn.it\\
	Galluccio@axpna1.na.infn.it\\
	Manko@axpna1.na.infn.it\\
	Miele@axpna1.na.infn.it
\end{itemize}}
\newpage

\section{Introduction}

A satisfactory description for the transverse dynamics of a charged particle
beam is a fundamental requirement for the new generation of accelerating
machines working at very high luminosity. To this aim, the recently proposed
{\it Thermal Wave Model\/} (TWM) \cite{fm1} could represent an interesting
framework for better describing the dynamics of charge particle beams.

According to this model, the beam transport is described in terms of a complex
function, the {\it beam wave function} (BWF), whose squared modulus gives the
transverse density profile. This function satisfies a Schr\"{o}dinger-like
equation in which Planck's constant is replaced by the transverse emittance.
In this equation, the potential term accounts for the total interaction
between the beam and the surroundings. In particular, in a particle accelerator
the potential has to take into account both the multipole-like contributions,
depending only on the machine parameters, and the collective terms,
which on the contrary depend on the particle distribution (self-interaction).

In transverse dynamics, TWM has been successfully applied to a number of linear
and non-linear problems. In particular, it seems to be capable of reproducing
the main results of Gaussian particle beam optics (dynamics for a
quadrupole-like device \cite{fm1}), as well as of estimating the luminosity in
final focusing stages of linear colliders in the presence of small sextupole-
and octupole-like deviations \cite{fm2}. In addition, for the case of
transverse dynamics in quadrupole-like devices with small sextupole and
octupole deviations, TWM predictions have been recently compared with tracking
code simulations showing a very satisfactory agreement \cite{fgm}.

More recently, in the framework of Nelson's stochastic mechanics the
Schr\"{o}dinger-like equation of TWM has been derived \cite{tzenov}.

This paper concerns with the transverse phase-space behaviour of a charged
particle beam passing through a thin quadrupole with small sextupole and
octupole deviations, in order to find  the most suitable function to describe
the {\it phase-space distribution} associated with the beam  in the framework
of TWM.

By using the BWF found in Ref.~\cite{fgm}, we compute the corresponding Wigner
function (WF) \cite{wig} which, according to the quantum mechanics formalism,
should represent the most natural definition for the beam distribution
function. Unfortunately, due to the uncertainty principle, the WF is not
positive definite, hence for the  above BWF we compute also the corresponding
Husimi function (HF) \cite{hus}, the so-called $Q$-function, which is
alternative to WF and is in fact positive definite. A comparison of both these
predictions with the results of a particle tracking simulation is then
performed in order to select the most appropriate definition of the beam
phase-space function in TWM.

The paper is organized as follows. A brief presentation of BWF determination is
given in section 2, whilst in section 3 we give the definition of both WF and
HF. In section 4, the comparison of the theoretical predictions with the
simulations results is presented. Finally, section 5 contains conclusions and
remarks.

\section{Beam Wave Function in presence of small
sextupole and octupole aberrations}

Let us consider a charged particle beam travelling along the $z$-axis
with velocity $\beta c$ ($\beta \approx 1$), and having transverse emittance
$\epsilon$.\\ We suppose that, at $z=0$, the beam enters a focusing
quadrupole-like lens of length $l$ with small sextupole and octupole
deviations, then it propagates in vacuo. In
this region, if we denote with $x$ the transverse coordinate (1-D case),
the beam particles feel the following potential
\begin{equation}
U(x,z) = \left\{ \begin{array}{cc}
\frac{1}{2!} k_{1} x^2 + \frac{1}{3!} k_{2} x^3 +\frac{1}{4!} k_{3} x^{4} &
0\leq z \leq l\\
0 &  z > l \end{array} \right.~~~,
\label{1}
\end{equation}
where $k_{1}$ is the quadrupole strength, $k_{2}$ is the sextupole strength and
$k_{3}$ is the octupole strength, respectively. Note that $U(x,z)$ is a
dimensionless energy potential, obtained dividing the potential energy
associated with the transverse particle motion
by the factor $m_{0}\gamma\beta^2c^2$, where $m_{0}$ and $\gamma$ are
the particle rest mass and the relativistic factor $[1-\beta^2]^{-1/2}$,
respectively.

As already stated, in the TWM, the transverse beam dynamics is ruled by the
Schr\"{o}dinger-like equation \cite{fm1}
\begin{equation}
i\epsilon~\frac{\partial \Psi}{\partial z}  = -
{}~\frac{\epsilon^2}{2} \frac{{\partial}^{2}}{\partial x^2} \Psi + U(x,z)
\Psi~~~.
\label{eq:schr}
\end{equation}
The $z$-constancy of the integral $\int_{-\infty}^{+\infty}
|\Psi(x,z)|^2~dz$, which is a consequence of the reality of $U(x,z)$
in (\ref{eq:schr}), suggests to interpret $|\Psi(x,z)|^2$
as the transverse density profile of the beam. Hence, if $N$ is the total
number of the beam particles, $\lambda (x,z)\equiv N~|\Psi (x,z)|^2$
is the transverse number density.

We fix the initial profile as a Gaussian density distribution of
r.m.s. $\sigma_{0}$, which corresponds to the initial BWF
\bee
\Psi_{0}(x)\equiv \Psi(x,0)=
{1 \over \left[2\pi\sigma_{0}^{2}\right]^{1/4}}
\exp\left(-\frac{x^2}{4\sigma^2_{0}}\right)~~~.
\label{eq:psiinit}
\ene
Provided that $\sigma_0 k_2/(3k_1) \ll 1$ and $\sigma_0^2 k_3/(12k_1)
\ll 1$, the quantum mechanics formalism for {\it time}-dependent perturbation
theory applied to (\ref{eq:schr})
for the case of a thin lens ($\sqrt{k_{1}}l \ll 1$)
allows us to give an approximate first-order normalized BWF in the
configuration space. At the exit of the lens ($z=l$) it reads
\cite{fgm}
\begin{eqnarray}
\Psi (x,l) & = & \Psi_{0}(x) ~\exp\left(-i {
K_1 x^2 \over 2 \epsilon} \right)
\left[ (1-i3\omega)~H_{0}\left(\frac{x}
{\sqrt{2}\sigma_{0}}\right) - i \frac{3\tau}{\sqrt{2}}~H_{1}\left(\frac{x}
{\sqrt{2}\sigma_{0}}\right)\right.
\nonumber\\
& - & \left. i 3 \omega ~H_{2}\left( \frac{x}{\sqrt{2}\sigma_{0}}\right)
- i \frac{\tau}{2\sqrt{2}}~H_{3}\left(\frac{x}{\sqrt{2}\sigma_{0}}\right)
-i
\frac{\omega}{4}~H_{4}\left(\frac{x}{\sqrt{2}\sigma_{0}}\right)\right]~~~,
\label{eq:psil}
\end{eqnarray}
where $\tau \equiv \sigma_0^3 K_2 /6\epsilon$, and $\omega \equiv \sigma_0^4
K_3 / 24\epsilon$, with $K_{i} \equiv k_{i} l$ ($i=1,2,3$), the integrated {\it
aberration strengths}. Remarkably, Eq. (\ref{eq:psil}) shows that, due to the
aberrations, the BWF after passing the {\it kick stage} is a superposition of
only five modes, according to the simple selection rules due to (\ref{1}).
Thus, after a drift of length $L$ ($L\gg l$) in the free space we get (see also
\cite{fgm})
\begin{eqnarray}
\Psi(x,L) & = & \frac{\exp{\left[- \frac{x^2}{4 \sigma^2(L)}\right]}
\exp{\left[i\frac{x^2}{2 \epsilon R(L)} + i \phi(L)\right]}}
{\left[2 \pi \sigma^2(L) (1 + 15 \tau^2 + 105 \omega^2)^2 \right]^{1/4}}
\left[ (1-i3\omega)~H_{0}\left(\frac{x}
{\sqrt{2}\sigma(L)}\right) \right.
\nonumber\\
& - & \left. i \frac{3\tau}{\sqrt{2}}~H_{1}\left(\frac{x}
{\sqrt{2}\sigma(L)}\right) e^{i 2 \phi(L)}
 - i 3 \omega ~H_{2}\left( \frac{x}{\sqrt{2}\sigma(L)}\right)
e^{i 4 \phi(L)} \right.
\nonumber\\
& - & \left. i \frac{\tau}{2\sqrt{2}}~H_{3}\left(\frac{x}{\sqrt{2}
\sigma(L)}\right) e^{i 6 \phi(L)}
- i \frac{\omega}{4}~H_{4}\left(\frac{x}{\sqrt{2}\sigma(L)}\right)
e^{i 8 \phi(L)}\right]~~~,
\label{13p}
\end{eqnarray}
where
\begin{eqnarray}
\sigma(L) & = & \left[\left( \frac{\epsilon^2}{4 \sigma_{0}^2} + K_{1}^2
\sigma_0^2\right) (L-l)^2 - 2 K_1 \sigma^2_0 (L-l) +
\sigma_0^2\right]^{1/2}~~~,
\nonumber\\
\frac{1}{R(L)} & = & \left.\frac{1}{\sigma}
\frac{d \sigma}{d z}\right|_{z=L}~~~,
\nonumber\\
\phi(L)& = & - \frac{1}{2} \left\{ \arctan{\left[
\left( \frac{\epsilon}{2 \sigma_{0}^{2}} + \frac{ 2 K_{1}^{2} \sigma^2_{0}}
{\epsilon}\right)\left( L-l \right)- \frac{ 2 K_{1} \sigma^2_{0}}
{\epsilon} \right]}\right.
\nonumber\\
& + & \left.\arctan{\left[ \frac{ 2 K_{1} \sigma^2_{0}} {\epsilon}\right]}
\right\}~~~.
\label{13q}
\end{eqnarray}
Correspondingly, the Fourier transform of $\Psi (x,z)$
\begin{equation}
\Phi(p,z)\equiv \frac{1}{\sqrt{2\pi}}\int_{-\infty}^{\infty}\Psi
(x,z)~\exp(-ipx/\epsilon)~dx~~~,
\label{3}
\end{equation}
is the BWF in the momentum space.\\
Consequently, $|\Psi (x,l)|^2$ and $|\Psi (x,L)|^2$ give the transverse
particle density profiles at $z=l$ (after lens) and at $z=L$ (after
drift), respectively, whilst  $|\Phi (p,l)|^2$ and $|\Phi (p,L)|^2$ give
the momentum distributions of the particles, at $z=l$ and at $z=L$,
respectively.

In Ref.~\cite{fgm} the configuration-space and momentum-space
distributions are reported for some significative values of parameters
$\sigma_{0}$, $\epsilon$, $K_{1}$, $K_{2}$, $K_{3}$, and $L$.
They are compared with the corresponding results obtained from
particle tracking simulations showing a very
satisfactory agreement.

In the next section we extend this analysis, made separately both in
configuration space and in momentum space, by introducing the
appropriate distribution function in the context of TWM approach.

Since the approximate solution (\ref{eq:psil}) and (\ref{13p}) for BWF
is given for small
sextupole and octupole deviations from a quadrupole potential (harmonic
oscillator), the present analysis falls in the {\it semiclassical} description
of the wave packet evolution (WKB theory) extensively treated in Ref.
\cite{littlejohn}.

\section{Phase-space distributions}

According to Quantum Mechanics (QM), for a given BWF $\Psi (x,z)$ we can
introduce the density matrix  $\rho$ as
\begin{equation}
\rho (x,y,z) \equiv \Psi (x,z) \Psi^{*}(y,z)~~~,
\label{eq:rhopsi}
\end{equation}
which, in the $\langle bra|$ and $|ket\rangle$ Dirac's notation,
is associated with the the following {\it density operator}
\begin{equation}
\hat{\rho} = | \Psi \rangle \langle \Psi |~~~.
\label{eq:matrdens}
\end{equation}
Note that $\hat{\rho}$ has the following two properties\\
i) probability conservation
\begin{equation}
\mbox{Tr}(\hat{\rho}) = 1~~~;
\label{eq:trace}
\end{equation}
ii) hermiticity
\begin{equation}
\hat{\rho}^{\dag} = \hat{\rho}~~~.
\label{eq:herm}
\end{equation}
On the basis of this density matrix definition, we can define the relevant
phase-space distributions associated with the transverse beam motion
within the framework of TWM.

\subsection{Wigner function} \label{sec:wigf}

One of the widely used phase-space representations given in QM is the
one introduced by Weyl and Wigner. In this representation,
by simply replacing Planck's constant with $\epsilon$, the phase-space
beam dynamics can be described in terms of the following function,
called Wigner function (WF)
\begin{equation}
W(x,p,z) \equiv { 1 \over 2 \pi \epsilon} \int_{-\infty}^{+\infty}
\rho\left(x-{y \over 2}, x+{y \over 2},z \right)~\exp\left(i {p y
\over\epsilon}
\right)~dy~~~,
\label{eq:wig}
\end{equation}
namely, by virtue of (\ref{eq:rhopsi})
\begin{equation}
W(x,p,z) = {1 \over 2 \pi \epsilon} \int_{-\infty}^{+\infty}
\Psi^{*}\left(x + { y \over 2},z\right) \Psi\left(x - { y \over 2},z\right)
{}~\exp\left(i{p y\over \epsilon} \right)~dy~~~.
\label{eq:wigexpr}
\end{equation}
It is easy to prove from (\ref{eq:wigexpr}) that
\begin{equation}
\int_{-\infty}^{+\infty}\int_{-\infty}^{+\infty}W(x,p,z)~dx~dp=1~~~,
\label{w-norm}
\end{equation}
\begin{equation}
\lambda (x, z)~=~N\int_{-\infty}^{+\infty} W(x,p,z)~dp~~~,
\label{eq:wigrho}
\end{equation}
and
\begin{equation}
\eta (p,z)~=~N\int_{-\infty}^{+\infty} W(x,p,z)~dx~~~,
\label{eq:wigintq}
\end{equation}
where $\eta (p,z)$ represents the transverse momentum space number density
(note that $\eta (p,z)=N ~ |\Phi (p,z)|^2$). Eqs. (\ref{eq:wigrho}) and
(\ref{eq:wigintq}) show that $N$ times $W(x,p,z)$ is the phase-space
distribution function associated with the transverse beam motion. Consequently,
$\lambda$ and $\eta$ are its configuration-space and momentum-space
projections, respectively.

The averaged-quantity description can be done by introducing the following
second-order momenta of $W$
\begin{equation}
\sigma_{x}^{2}(z) \equiv
\int_{-\infty}^{+\infty}\int_{-\infty}^{+\infty}x^2~W(x,p,z)~dx~dp
\equiv \langle x^2 \rangle~~~,
\label{eq:sigmax2}
\end{equation}
\begin{equation}
\sigma_{xp}(z) \equiv
\int_{-\infty}^{+\infty}\int_{-\infty}^{+\infty}xp~W(x,p,z)~dx~dp
\equiv {1 \over 2}\langle xp+px \rangle~~~,
\label{eq:sigmaxp}
\end{equation}
\begin{equation}
\sigma_{p}^{2}(z) \equiv
\int_{-\infty}^{+\infty}\int_{-\infty}^{+\infty}p^2~W(x,p,z)~dx~dp
\equiv \langle p^2 \rangle~~~.
\label{eq:sigmap2}
\end{equation}
They are connected with the geometrical properties of the phase-space ensemble
associated with the beam; here we have considered the case $\langle p
\rangle=\langle x \rangle=0$.

According to QM and simply replacing $\hbar$ with $\epsilon$, we
can easily say that $W(x,p,z)$, defined by (\ref{eq:wigexpr})
when the BWF $\Psi (x,z)$ is
solution of (\ref{eq:schr}) (with $U$ arbitrary Hermitian potential),
satisfies the following Von Neumann equation
\begin{equation}
\left[{\partial \over \partial z}~+~p{\partial \over \partial x}~+~{i \over
\epsilon}\left(U\left(x+{i\epsilon \over 2}{\partial \over \partial
p}\right)~-~U\left(x-{i\epsilon \over 2}{\partial \over \partial
p}\right)\right)\right]W~=~0~~~.
\label{von-neumann}
\end{equation}

Thus, in order to perform a phase-space analysis, one has two
possibilities: either
to solve (\ref{eq:schr}) for $\Psi (x,z)$ and then, by means of the Wigner
transform (\ref{eq:wigexpr}), to obtain $W(x,p,z)$, or
to solve directly (\ref{von-neumann}) for $W(x,p,z)$.

Although for the potential given by (\ref{1}) the exact
solution of the Schr\"{o}dinger-like
equation (\ref{eq:schr}) is unknown, time-dependent
perturbation theory can be applied to give, at any order
of the perturbative expansion, the approximate solution \cite{landau}:
for example, (\ref{eq:psil}) is the first-order approximate solution
of (\ref{eq:schr}) for the potential (\ref{1}).\\
Also for the Von Neumann equation, (\ref{von-neumann}), the exact solution  is
not available in the case of potential (\ref{1}), with the exception of the
case with $k_2=k_3=0$ (pure quadrupole/harmonic
oscillator).
Nevertheless, it has been shown  \cite{narcowich} that,
for a more general Hamiltonian which
includes our case as a special case,
a perturbative Dyson-like expansion
for the Wigner function can be constructed which
converges to the solution of the corresponding quantum Liouville equation.
In general this approach, providing  $W(x,p,z)$
directly from the phase-space dynamics, should yield a smaller error.

For a thin lens the Wigner transform, (\ref{eq:wigexpr}), of the approximate
BWF, as given by (\ref{eq:psil}) and (\ref{13p}), coincides with the
approximate
first-order  solution of quantum Liouville equation \cite{narcowich}:
the two approaches are therefore equivalent.
As TWM has been mainly developed in the configuration space \cite{fm1}, and
the approximate BWF for the typical potential used in particle
accelerators has already been calculated (see, for example,
\cite{fm1}-\cite{fgm}), in this paper we have chosen to proceed with this
latter approach.

For what concerns the calculation of the higher-order solutions,
solving directly  (\ref{von-neumann}) for $W(x,p,z)$
should, in principle, be preferable
to the procedure in which we first solve for BWF and then use
(\ref{eq:wigexpr}), because less approximations are required, and therefore
smaller errors are involved.
Unfortunately it is very difficult to treat the Von Neumann equation,
(\ref{von-neumann}), numerically,
especially if the classical potential contains powers in $x$ higher
than the quadratic one. In fact, in this case operators
$U\left[x \pm i(\epsilon/2){\partial \over \partial p}\right]$ make
(\ref{von-neumann}) a partial differential equation of order
higher than the second in the $p$-derivative, which is rather
difficult to handle numerically.
In order to treat BWF and Wigner transform numerically, instead, one
can profit of the very powerful methods developed for the
Schr\"odinger equation, as it has been done, for instance, in Ref. \cite{fmve}.

Although $W$ is the distribution function of the
system in the framework of TWM, due to well-known quantum mechanical
properties, it is not positive definite. However, in QM it results to be
positive for some special harmonic oscillator wave functions, called {\it
coherent states} \cite{coherstates,coherstates1,coherstates2,coherstates3},
which give purely Gaussian density profiles.

Coherent states for charged particle beams have been recently introduced in TWM
in order to describe the coherent structures of charged particle distributions
produced in an accelerating machine \cite{dnfmm}. The fact that $W$ can assume
negative values for some particular cases, reflects the quantum-like properties
of both the wave function and the density operator.

Remarkably, in the case of a pure quadrupole-like lens ($U=k_1 x^{2}/2$)
an interesting quantity, which
estimates the r.m.s. area of the ensemble, can be expressed in terms of these
momenta, namely
\begin{equation}
\pi\left[ \langle x^2 \rangle ~ \langle p^2 \rangle
- {1 \over 4}\langle xp+px \rangle^2\right]^{1/2} =
\pi\left[\sigma^2_{x}(z) \sigma^2_{p}(z) - \sigma^2_{xp}(z) \right]^{1/2}~~~.
\label{eq:area}
\end{equation}
Since in this case the ground-like state is
\bee
\Psi_{0}(x,z)=
{1 \over \left[2\pi\sigma_{0}^{2}\right]^{1/4}}
\exp\left(-\frac{x^2}{4\sigma^2_{0}}\right)\exp\left(-\frac{i}{2}\sqrt{k_{1}}z
\right)~~~,
\label{eq:psigauss}
\ene
the formula (\ref{eq:wigexpr}) produces the following bi-Gaussian Wigner
distribution
\begin{equation}
W_{0}(x,p) = { 1\over \pi \epsilon} \exp\left( - { x^2 \over 2
\sigma_{0}^2} - { 2 \sigma_{0}^2 \over
\epsilon^2}p^2  \right)~~~.
\label{eq:wig0}
\end{equation}
In addition, for the following non-coherent Gaussian-like state associated
with the charged-particle beam \cite{dnfmm}
\begin{equation}
\Psi(x,z) = { 1 \over \left[ 2 \pi \sigma^2(z)\right]^{1/4}}
\exp\left[ - { x^2 \over 4 \sigma^2(z) }
+ i { x^2 \over 2 \epsilon R(z) }
+ i \phi(z) \right]~~~,
\label{eq:non-coherent}
\end{equation}
where
\begin{equation}
{d^2\sigma \over dz^2}+k_{1}\sigma-{ \epsilon^2\over 4\sigma^3}=0~~~~,~~~~
{1\over R}={1\over \sigma}{d\sigma\over dz}~~~~,~~~~
{d\phi\over dz}=-{\epsilon\over 4\sigma^2}~~~,
\label{eq:envelope}
\end{equation}
definition (\ref{eq:wigexpr}) of $W$ easily gives
\begin{equation}
W(x,p,z) = { 1 \over \pi \epsilon}
\exp\left[ - { x^2 \over 2 \sigma^2(z) }
- { 2 \sigma^2(z) \over \epsilon^2} \left( { x \over R(z)} - p
\right)^2 \right]~~~.
\label{eq:wigz}
\end{equation}
Note that in this simple case of pure quadrupole the exact analytical solution
(\ref{eq:wig0}) and (\ref{eq:wigz}) can be easily obtained
directly by integrating (\ref{von-neumann}).

Note also that the argument of the exponential in (\ref{eq:wigz}) is a
quadratic
form in the variables $x$ and $p$, which can be written as
\begin{equation}
F(x,p,z) \equiv - {2 \over \epsilon} \left[
\gamma(z) x^2 + 2 \alpha(z) x p + \beta(z) p^2 \right]~~~,
\label{eq:quadraticform}
\end{equation}
where
\begin{equation}
\alpha(z)=-{\sigma^2(z) \over \epsilon R(z)}~~~~,~~~~
\beta(z)={\sigma^2(z) \over \epsilon}~~~~,~~~~
\gamma(z) = {\epsilon \over 4 \sigma^2(z)} + {\sigma^2(z)\over
\epsilon R^2(z)}~~~.
\label{eq:twissdef}
\end{equation}
These quantities are usually called Twiss parameters \cite{lawson}.
By substituting (\ref{eq:wigz}) into (\ref{eq:sigmax2})-(\ref{eq:sigmap2}),
we obtain $\alpha(z)= - (1/2) d\beta(z)/dz$,
$\gamma(z) = \sigma_p^2(z)/\epsilon$ and
\begin{equation}
\sigma_x(z) =\sigma(z)~~~,~~~~~~~~\sigma_p^2(z) =
{\epsilon^2 \over 4 \sigma^2(z)} + {\sigma^2(z) \over R^2(z)}=
{\epsilon^2 \over 4 \sigma^2(z)} ( 1 + 4 \alpha^2(z))~~~.
\label{eq:sigmaxsigmap}
\end{equation}
Consequently, we immediately get the following quantum-like version of
the Lapostolle definition of emittance \cite{lawson}
\begin{equation}
\langle x^2 \rangle  \langle p^2\rangle  - {1 \over 4}
\langle xp+px \rangle^2
= {\epsilon^2 \over 4} = \mbox{const.}~~~,
\label{eq:courant}
\end{equation}
from which uncertainty relation of TWM can be easily derived
\begin{equation}
\sigma^2(z) \sigma_p^2(z) \geq { \epsilon^2 \over 4}~~~.
\label{eq:TWM-uncertainty}
\end{equation}
Furthermore, the equation
\begin{equation}
\gamma(z) x^2 + 2 \alpha(z) xp + \beta(z) p^2 ={\epsilon \over 2}~~~,
\label{eq:ellipse}
\end{equation}
represents, for each $z$, an ellipse in the phase-space of area $\pi\epsilon/2$
associated with the particle beam motion.

Consequently, the operator
\begin{equation}
\hat{{\cal J}} (x,p,z)~\equiv ~ \gamma (z) x^2~+~2\alpha (z) {xp+px \over
2}~+~\gamma (z) p^2
\label{courant-snyder-invariant}
\end{equation}
is the quantum-like version of one of the well-known Courant-Snyder invariant
\cite{courant-snyder}. In fact, it easy to prove that
\begin{equation}
i\epsilon{\partial\hat{{\cal J}}\over\partial z}~+~
\left[\hat{{\cal J}}, \hat{H}\right] ~=~0~~~,
\label{quantum-like-total-derivative}
\end{equation}
where $\hat{H}$ is the Hamiltonian operator for the case of a quadrupole.

\subsection{Husimi function}

The phase-space description of a quantum system can be done in terms of
another function which has been introduced by Husimi \cite{hus}, the
so-called {\it $Q$-function} or {\it Husimi function} (HF).
In order to give an analogous definition of HF for
charged particle beams in the context of TWM, we still replace
Planck's constant with $\epsilon$, obtaining
\begin{equation}
Q(x_{0},p_{0},z)\equiv {1 \over
2\pi\epsilon}~\int_{-\infty}^{\infty}\int_{-\infty}^{\infty}~\Theta^{*}
(u,z;x_{0},p_{0}) \rho (u,v,z)\Theta (v,z;x_{0},p_{0})~du~dv~~~,
\label{eq:q-fun}
\end{equation}
where $\Theta (x,z;x_{0},p_{0})$ is a coherent state associated with the
charged particle beam defined as \cite{dnfmm}
\bee
\Theta (u,z;x_{0},p_{0})=
{1 \over \left[2\pi\sigma_{0}^{2}\right]^{1/4}}
\exp\left[-\frac{(u-x_{0}(z))^2}{4\sigma^2_{0}}+\frac{i}{\epsilon}p_{0}(z)u
-i\delta_{0}(z)\right]~~~,
\label{eq:thetacoher}
\ene
with
\bee
x_{0}(z)\equiv \langle u \rangle =
\int_{-\infty}^{\infty}u~|\Theta|^2~du~~~,
\label{xaverage}
\ene
\bee
p_{0}(z)\equiv \langle \hat{p} \rangle =
\int_{-\infty}^{\infty}\Theta^{*}~
\left(-i\epsilon{\partial \over \partial x} \right)\Theta~dx~~~,
\label{paverage}
\ene
and
\bee
{d\delta_{0} \over dz}= {p_{0}^2 \over 2\epsilon}-\sqrt{k_{1}}{x_{0}^2 \over
4\sigma_{0}^2}+{\sqrt{k_{1}}\over 2}~~~.
\label{deltazero}
\ene
Note that here $x_{0}$ and $p_{0}$ play the role of classical phase-space
variables.\\
By substituting (\ref{eq:rhopsi}) and (\ref{eq:thetacoher}) in
(\ref{eq:q-fun}) we obtain
\begin{eqnarray}
Q(x,p,z)& = & {1 \over 2\pi\epsilon \sqrt{2\pi\sigma_{0}^{2}} }
\exp\left(-{x^2 \over 2 \sigma_{0}^{2}}\right)
\int_{-\infty}^{\infty}dv\int_{-\infty}^{\infty}du
\nonumber\\
& \times &\exp\left[-{u^2+v^2 \over
4\sigma_{0}^{2}}+{x(u+v)\over 2\sigma_{0}^{2}}+{i\over\epsilon}p(u-v)\right]
\Psi^{*}(u,z) \Psi(v,z)~~~,
\label{eq:q-xpz}
\end{eqnarray}
where for simplicity we have replaced the classical variables $x_{0}$ and
$p_{0}$ with $x$ and $p$, respectively.

However, the integration over $p$ and over
$x$, does not give configuration-space and momentum-space
distributions, respectively. But this function removes the {\it pathology} of
negativity exhibited by $W$, and thus gives a more accurate description in the
phase-space region where $W$ is negative.\\ Some properties of $Q$ are in
order.

\vspace{.5cm}

\noindent
i) It is easy to prove the following normalization relation
\begin{equation}
\int_{-\infty}^{\infty}\int_{-\infty}^{\infty}Q(x,p,z)~dx~dp=1~~~.
\label{eq:Q-norm}
\end{equation}

\vspace{.5cm}

\noindent
ii) Definition (\ref{eq:q-xpz}) can be cast in the following more convenient
form
\begin{equation}
Q(x,p,z) = { 1 \over 8 \pi \epsilon \sqrt{2 \pi \sigma_0^2}}
\left| \int_{-\infty}^{\infty}dy~\exp\left( - { y^2 \over 16 \sigma_0^2}
- i {p y \over 2 \epsilon} \right)
{}~\Psi\left(x + { y \over 2},z \right)
 \right|^2~~~.
\label{eq:q-convform}
\end{equation}
Note that (\ref{eq:q-convform}) clearly shows that $Q$ is positive definite.

\vspace{.5cm}

\noindent
iii) Definition (\ref{eq:q-fun}) of Husimi function, or equivalently
(\ref{eq:q-convform}), does not give in general a phase-space distribution
coinciding with WF. But, in the case of a Gaussian BWF
they must coincide, because in this case both $Q$ and $W$ reduce to
the {\it classical} distribution. In particular, we observe that
since in (\ref{eq:q-fun}) or in (\ref{eq:q-convform}) the constant $\sigma_0$
is involved, the present definitions of $Q$ are suitable only to describe
the phase-space distribution of eigenstates: this way $Q$ does not explicitly
depend on $z$. The natural generalization of $Q$ for $z$-dependent BWF
will be introduced later. Now we point out that, in connection with
the Gaussian BWF given by (\ref{eq:psigauss}), Eq. (\ref{eq:q-convform})
gives a bi-Gaussian $Q$-function that does not coincide with (\ref{eq:wig0})
because of a scaling disagreement. This problem can be easily removed, by
introducing the following new definition for $Q$
\begin{equation}
Q(x,p,z) = { 1 +\lambda^2 \over 8 \pi \epsilon \lambda^2
\sqrt{2 \pi \sigma_0^2}}
\left| \int_{-\infty}^{\infty}dy~\exp\left( - { y^2 \over 16 \sigma_0^2
\lambda^2} - i { \sqrt{1 +\lambda^2} p y \over 2 \epsilon \lambda } \right)
{}~\Psi\left(\sqrt{1 + \lambda^2} x + { y \over 2},z \right)
 \right|^2,
\label{eq:q-lambdaform}
\end{equation}
for any real number $\lambda$. This is equivalent to introduce the following
substitutions in the definition (\ref{eq:q-fun})
\begin{equation}
x \rightarrow \sqrt{1+\lambda^2}~x~~~, ~~~p \rightarrow {\sqrt{1+\lambda^2}
\over \lambda}~p~~~, ~~~Q\rightarrow {1+\lambda^2 \over \lambda}~Q~~~.
\label{substitution}
\end{equation}
Under these substitutions, the normalization condition (\ref{eq:Q-norm}) is
preserved. In order to symmetrize (\ref{substitution}),
we choose $\lambda=1$. Consequently,
\begin{equation}
x \rightarrow \sqrt{2}~x~~~, ~~~p \rightarrow \sqrt{2}~p~~~,
{}~~~Q\rightarrow 2~Q~~~,
\label{substitution1}
\end{equation}
and (\ref{eq:q-lambdaform}) becomes
\begin{equation}
Q(x,p,z) = { 1  \over 4 \pi \epsilon \sqrt{2 \pi \sigma_0^2}}
\left| \int_{-\infty}^{\infty}dy~\exp\left( - { y^2 \over 16 \sigma_0^2 }
- {i   \over \sqrt{2} \epsilon }p y \right)
{}~\Psi\left(\sqrt{2} x + { y \over 2},z \right)
 \right|^2~.
\label{eq:q-fun1}
\end{equation}
Thus, by using (\ref{eq:psigauss}) in (\ref{eq:q-fun1}) we obtain
\begin{equation}
Q(x,p) = { 1\over \pi \epsilon} \exp\left( - { x^2 \over 2
\sigma_{0}^2} - { 2 \sigma_{0}^2 \over
\epsilon^2}p^2  \right)~~~,
\label{eq:q0}
\end{equation}
which now coincides with the corresponding Wigner function
(\ref{eq:wig0}).

\vspace{.5cm}

\noindent
iv) For the non-coherent Gaussian-like BWF (\ref{eq:non-coherent}), Eq.
(\ref{eq:q-fun1}) is no longer valid, because it does not coincide with
(\ref{eq:wigz}). In particular, at any $z$, the phase-space ellipses associated
with $W$ do not coincide with the contours of (\ref{eq:q-fun1}) for BWF given
by
(\ref{eq:non-coherent}). This problem can be overcome by generalizing
definition (\ref{eq:q-fun1}) with the following
\begin{equation}
Q(x,p,z) = { 1  \over 4 \pi \epsilon \sqrt{2 \pi \sigma^2(z)}}
\left| \int_{-\infty}^{\infty}dy~\exp\left( - { y^2 \over 16
\sigma^2(z) } - { i y^2 \over 8 \epsilon R(z)}
- {i   \over \sqrt{2} \epsilon }p y \right)
{}~\Psi\left(\sqrt{2} x + { y \over 2},z \right)
 \right|^2~~~.
\label{eq:q-fun2}
\end{equation}
By substituting (\ref{eq:non-coherent}) in (\ref{eq:q-fun2}) we easily obtain
\begin{equation}
Q(x,p,z) = { 1 \over \pi \epsilon}
\exp\left[ - { x^2 \over 2 \sigma^2(z) }
- { 2 \sigma^2(z) \over \epsilon^2} \left( { x \over R(z)} - p
\right)^2 \right]~~~,
\label{eq:qz}
\end{equation}
which now coincides with the corresponding WF given by (\ref{eq:wigz}).

\vspace{.5cm}

(v) Note that the generalized definition (\ref{eq:q-fun2}) suggests to start
from (\ref{eq:q-fun}) where the coherent states $\Theta (u,z;x_0,p_0)$ are
replaced by the following non-coherent Gaussian-like states
\bee
\Theta_0 (u,z;x_{0},p_{0})=
{1 \over \left[2\pi\sigma^{2}(z)\right]^{1/4}}
\exp\left[-\frac{(u-x_{0}(z))^2}{4\sigma^2(z)}+\frac{i}{\epsilon}p_{0}(z)u
-i\delta_{0}(z) + \frac{(u-x_{0}(z))^2}{2 \epsilon R(z)}\right]~~~.
\label{eq:thetaz}
\ene
This way if we start from (\ref{eq:q-fun}) with (\ref{eq:thetaz}), it is very
easy to prove that it is equivalent to (\ref{eq:q-fun2}) for any $\Psi$.
Consequently, the generalized definition of (\ref{eq:q-fun}) with
(\ref{eq:thetaz}) produce the same results of Wigner function definition
(\ref{eq:wigexpr}) for any Gaussian-like states.

\vspace{.5cm}

\noindent
vi) Note that (\ref{eq:thetaz}) are the non-coherent fundamental
modes of the following set of solutions of (\ref{eq:schr}) in the case
of $U = (1/2) K_1 x^2$
\begin{equation}
\Theta_n(x,z;x_0,p_0) = {\Theta_0(x,z;x_0,p_0) \over \sqrt{2^n n!}}
H_n\left( { x - x_0(z)\over \sqrt{2} \sigma(z) }\right)~\exp\left(
i 2 n \phi(z)\right)~~~.
\label{eq:theta-n}
\end{equation}
These modes are analogous to squeezed and correlated Fock's states used in
quantum optics \cite{Dodonov}.
In the next section we check if the equivalence between (\ref{eq:q-fun})
with (\ref{eq:thetaz}) and (\ref{eq:wigexpr}) is still valid for
non-Gaussian BWF. In particular, we make a comparison between
definition (\ref{eq:wigexpr}) and (\ref{eq:q-fun}) with
(\ref{eq:thetaz}) for the case of BWF given by (\ref{eq:psil}) and (\ref{13p}),
and in addition we compare them with the corresponding results of a
multi-particle tracking simulation.

\section{Numerical analysis}

A numerical study has been pursued in order to compare the description of
the phase-space as given by TWM, both by means of WF and by means of HF,
with the one resulting from standard particle tracking.

A Gaussian flat (1-D) particle beam has been used as starting beam, with
emittance $\epsilon = 120 \times 10^{-6} \mbox{m}~ \mbox{rad}$, and $\sigma_0 =
 0.05$ m. From Eq.~(\ref{eq:sigmaxsigmap}) it follows that, if $\alpha(0) = 0$
at the start, $\sigma_{p0}\equiv \sigma_{p}(0) =  1.2 \times 10^{-3}~\mbox{
rad}$. (Note that the definition of emittance given in (\ref{eq:courant})
differs from the definition used in classical accelerator physics by a factor
1/2.)

A simple device made of a quadrupole magnet  plus a drift space has been
considered as beam transport line: in addition sextupole and/or  octupole
aberrations have been included in the quadrupole.

The Wigner function (\ref{eq:wigexpr}) and the $Q$-function (\ref{eq:q-fun2})
have been computed by numerical integration  for different combinations of
aberration strengths, and have been compared with the results of the tracking
of $7 \times 10^{5}$ particles.

Isodensity contours at $1$, $2$ and $3~\sigma$ have been used to describe the
particle distribution in phase-space, both before and after the passage through
the simple device specified above. It is worth noting that with this choice
only $2\%$ of the particles are found beyond the contour at $2~\sigma$, and
only $0.01\%$ of them are beyond the contour at $3~\sigma$.

In Figure~1. the starting distribution is shown together with the
distribution emerging from the linear transport line. As already shown, in the
linear case both WF and HF can be computed analytically (Eqs.~(\ref{eq:wigz})
and (\ref{eq:qz})) yielding exactly the same result of the conventional
accelerator physics. Therefore, the output from tracking
and the TWM predictions are in full agreement, and thus their superposition is
not shown here.

In the cases shown in Figure~2. non-linear perturbations have been
added to the quadrupole lens: a sextupole perturbation in the first column, an
octupole perturbation in the second column, and perturbations of both kinds in
the third column. The results from the numerical computation of WF and HF are
shown also superposed to the tracking results for a first set of perturbation
values, corresponding to the distortions\footnote{ The phase-space distortion
can be defined as the ratio between the deflection $\Delta p(x)\equiv p(x) -
p(x_0)$, at $x=\sigma_0$, and $\sigma_{p0}$, namely $D\equiv \Delta
p(\sigma_0)/\sigma_{p0}$. It is easy to prove that $D=6 \tau= \sigma_0^3 K_2
/\epsilon$ for pure sextupole, and $D= 8 \omega = \sigma_0^4 K_3 / 3 \epsilon$
for pure octupole.} $D=0.125$  in the case of $K_2 = 0.06~m^{-2}$ and $D =
0.042$ in the case of $K_3 = 1.2~m^{-3}$. It should be noted that these values
of distortion, chosen with the aim of enhancing the effects desired, are huge
compared with what could be considered acceptable in any realistic single pass
device.

A quite good agreement with tracking can be observed for the contours at $1$
and  $2~\sigma$ for both WF and HF. The contours at $3~\sigma$ show some
discrepancies, more pronounced in the case of the Wigner function, which, as
discussed in Section~\ref{sec:wigf}, in the periphery of the distribution
produces regions with negative phase-space density which yield an  unrealistic
distortion of the phase-space. By the use of the $Q$-function, instead, this
effect is largely smoothed out, as shown in the last two raws of
Figure~2..

In the plots  displayed in the different columns of Figure~3. the same
kinds of perturbations are included as in Figure~2., but with
twice the strength, therefore yielding twice the distortion.

These large values of distortion approach the limits of applicability of
perturbation theory which, for the starting parameters and the quadrupole
strength selected, are given by $D \ll 3{\sigma_0}^2 K_1 / \epsilon =
2.25$ for the sextupole perturbation and $D \ll 4{\sigma_0}^2 K_1 /
\epsilon = 3.00$ for the octupole perturbation.

Indeed larger discrepancies between TWM and tracking can be observed now, in
particular where the sextupole perturbation is present ($D~=~0.25$).
Nevertheless in these cases (columns 1 and 3 of Figure~3.) HF still
describes fairly well the phase-space distribution  for particles with
amplitudes up to $1-1.5~\sigma$, whilst the WF distortion due to its negativity
starts to show up already at amplitudes of the order of 1$\sigma$.

In spite of the discrepancies observed at large amplitudes due to the
particularly strong perturbations used in this study, these results can be
considered very satisfactory: the TWM phase-space description of beam dynamics
in the presence of a non-linear lens, and in particular the one given by the
HF, will be in more than reasonable agreement with the one given by classical
accelerator physics for all realistic values of perturbation.

\section{Remarks and conclusions}

In this paper we have studied the phase-space distribution associated with the
transverse motion of a charged particle beam travelling through a
quadrupole-like device with sextupole and octupole deviations. In particular, a
quantum-like phase-space analysis within the {\it Thermal Wave Model} has been
developed and its predictions have been compared with the results of particle
tracking simulations.

To this end, we have first introduced the density matrix for the beam wave
function (\ref{eq:rhopsi}). Then, following the usual definitions, we have
constructed the Wigner transform (Eq.~(\ref{eq:wigexpr})), and the Husimi
transform
(Eq.~(\ref{eq:q-fun})) with $\Theta$ given by (\ref{eq:thetaz}). These
functions represent the best candidates for describing the full
phase-space evolution of the beam.
Our aim was to compare the results of tracking
simulations with the predictions of TWM, given in terms of $W$ and $Q$, in
order to enquire if the equivalence between (\ref{eq:q-fun})
with (\ref{eq:thetaz}) and (\ref{eq:wigexpr}) is valid also for
non-Gaussian BWF, and thus to
determine the appropriate function to describe the phase-space
dynamics.

The Wigner transform of the BWF, which represents the {\it natural} quantum
analogous of the classical phase-space distribution associated with the beam,
has
been numerically computed for the present problem. Its $x$ and $p$-projections
reproduce the configuration and momentum space distribution well \cite{fgm}, up
to the first order of the perturbation theory. According to the results
presented in section 4, this function reveals to be in good agreement with the
tracking results for small values of the integrated multipole-like strengths,
for which conditions $\sigma_0 K_2/(3K_1) \ll 1$ and $\sigma_0^2 K_3/(12K_1)
\ll 1$ are fully satisfied. For larger values of these parameters a little
discrepancy appears. The reason is that in the last case the first-order
perturbative expansion is not reliable enough. Nevertheless, we stress the
fact that, beyond the contours at $2~\sigma$ and $3~\sigma$, only the $2\%$ and
$0.01\%$ of the beam particles are present, respectively.
Summarizing, the comparison between the tracking results and the TWM
predictions shows that:\\ i) for small aberrations (conditions $\sigma_0
K_2/(3K_1) \ll 1$ and $\sigma_0^2 K_3/(12K_1) \ll 1$ well satisfied), both $Q$
and $W$ are in good agreement with the tracking results;\\ ii) for larger
aberrations, WF and HF exhibit the same order of discrepancy with respect to
the tracking contours, but the distortions in the Wigner contours are much more
evident due to the negativity of this function which is responsible for a
change in the contour concavity.

In conclusion, up to the first order in the perturbation theory, and for thin
lens approximation, it results that Wigner function can be adopted as
appropriate phase-space distribution in TWM. In fact it gives the correct $x$-
and $p$-projections, which are in good agreement with the tracking
configuration and momentum distribution, respectively. Finally, although Husimi
function does not give the right $x$ and $p$-projections, according to our
analysis it provides a better description as far as the only phase-space
dynamics is concerned.

\newpage

{\Large \bf Figure Captions}

\begin{itemize}
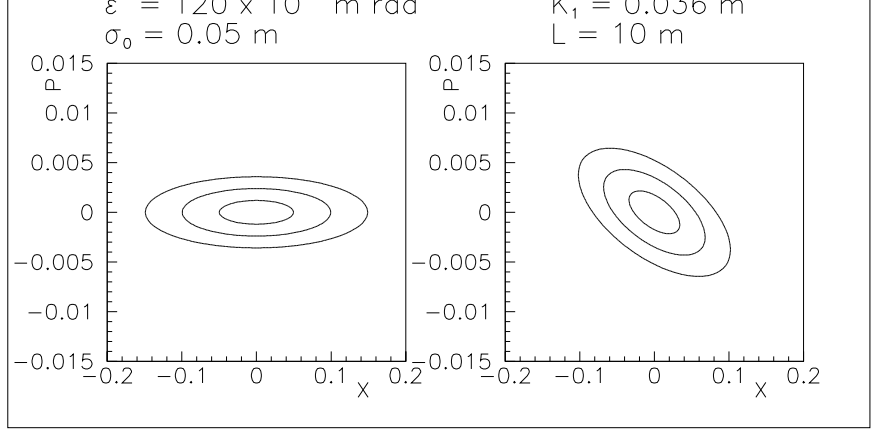
\item[Figure 1.]
Phase-space distribution at the starting point (left) and after a
quadrupole lens plus drift space (right). Starting from the center of the
distributions the isodensity contours correspond to $1$, $2$ and $3~\sigma$,
respectively.
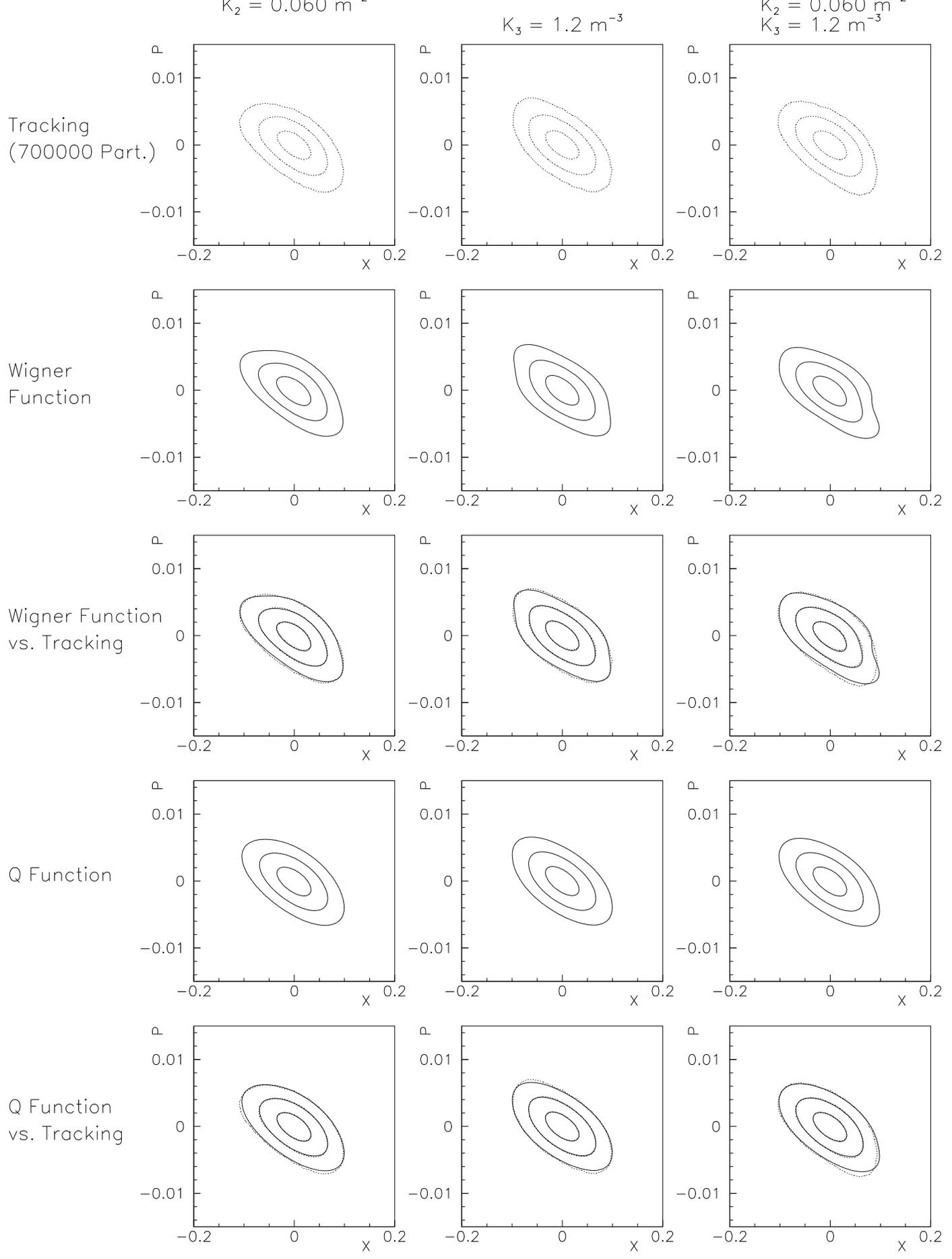
\item[Figure 2.]
Comparison of TWM phase-space description with tracking (dotted
lines) results. Parameter set No.~1. Starting from the center of the
distributions the isodensity contours correspond to $1$, $2$ and $3~\sigma$,
respectively.
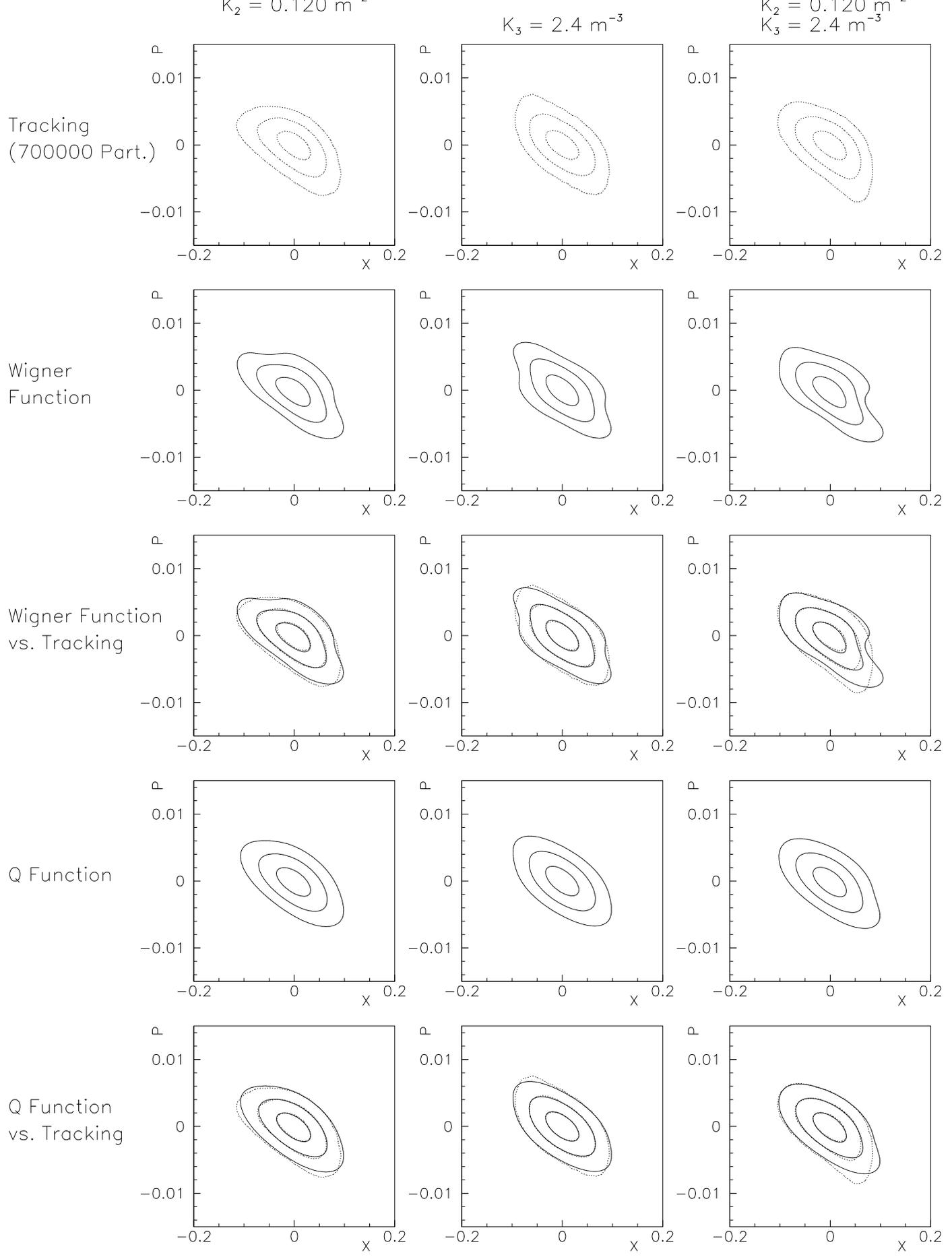
\item[Figure 3.]
Comparison of TWM phase-space description with tracking (dotted
lines) results. Parameter set No.~2. Starting from the center of the
distributions the isodensity contours correspond to $1$, $2$ and $3~\sigma$,
respectively.
\end{itemize}

\newpage

\begin{thebibliography}{99}
\setlength{\itemsep}{-2pt}

\bibitem{fm1} R. Fedele and G. Miele,  Nuovo Cimento D {\bf 13}, (1991) 1527.

\bibitem{fm2} R. Fedele and G. Miele, Phys. Rev. A {\bf 46}, (1992) 6634.

\bibitem{fgm} R. Fedele, F. Galluccio, and G. Miele, Phys. Lett. A {\bf 185},
(1994) 93; {\it A Numerical Check of the Thermal-Wave Model for Particle-Beam
Dynamics}, in Proc. of {\it Particle Accelerator Conference 93}, Washington
D.C., 17-20 May 1993 (IEEE) 209-211.

\bibitem{tzenov} S.I. Tzenov, {\it The Concept of Stochastic Mechanics in
Particle Accelerators}, to be published in Proc. of the Workshop on
{\it Nonlinear Dynamics in Particle Accelerators: Theory and Experiments},
september 5-9, 1994, Arcidosso, Italy (published by AIP).

S.I. Tzenov, {\it The Schr\"odinger Equation with Electromagnetic Potentials
in the Framework of Stochastic Quantization Approach}, INFN/TC-95/21, (1995).

\bibitem{wig} E. Wigner, Phys. Rev. {\bf 40}, 749 (1932).

\bibitem{hus} K. Husimi, Proc. Phys. Math. Soc. Japan {\bf 22}, 264 (1940).

\bibitem{littlejohn} R.G. Littlejohn, Phys. Rep. {\bf 138}, (1986) 193-291.

\bibitem{landau} L.D. Landau and E.M. Lifshitz, {\it Quantum Mechanics}
(Pergamon, London, 1958).

\bibitem{narcowich} F.J. Narcowich, J. Math. Phys. {\bf 10}, (1986) 2502.

\bibitem{fmve} R. Fedele, G. Miano, and L. Verolino, {\it An Investigation
of the Thermal-Wave Model Using the Galerkin Method}, INFN/TC-94/22, (1994).

\bibitem{coherstates} R. Glauber, Phys. Rev. Lett. {\bf 10}, 84 (1963).

\bibitem{coherstates1} E.C.G. Sudarshan, Phys. Rev. Lett. {\bf 10}, 277 (1963).

\bibitem{coherstates2} J.R. Klauder, J. Math. Phys. {\bf 5}, 177 (1964).

\bibitem{coherstates3} I.A. Malkin and V.I. Man'ko, {\it Dynamical
Symmetries and Coherent States of Quantum Systems} (in Russian),
(Nauka, Moscow, 1979).

\bibitem{dnfmm} S. De Nicola, R. Fedele, V.I. Man'ko, and G. Miele,
Physica Scripta {\bf 52}, (1995) 191.

\bibitem{lawson} J.D. Lawson {\it The Physics of Charged-Particle Beams},
2nd Edition (Clarendon, Oxford, 1988).

\bibitem{courant-snyder} E.D. Courant and H.S. Snyder, Ann. Phys.
{\bf 3}, (1958) 1.

\bibitem{Dodonov} V. V. Dodonov and V. I. Man'ko, {\it Invariants and
Evolution of Non-stationary Quantum Systems}, Proceedings of Lebedev
Physical Institute {\bf 183}, ed. M. A. Markov (Nova Science, Commack,
N. Y., 1989).
\end{thebibliography}
\end{document}